\documentclass[10pt]{article}
\usepackage[utf8]{inputenc}
\usepackage[english]{babel}
\usepackage[round]{natbib}
\usepackage{amsmath}
\usepackage{times}
\usepackage[title]{appendix}
\usepackage{amssymb}
\usepackage{ulem}
\usepackage{multicol,caption}
\usepackage{graphicx}
\usepackage{titlesec}
\usepackage{subfigure}
\usepackage{verbatim}
\usepackage{float}
\usepackage{titlesec}
\usepackage{hyperref}
\usepackage{multirow}
\usepackage{bm}
\usepackage{color}
\makeatletter
\def\UrlAlphabet{%
      \do\a\do\b\do\c\do\d\do\e\do\f\do\g\do\h\do\i\do\j%
      \do\k\do\l\do\m\do\n\do\o\do\p\do\q\do\r\do\s\do\t%
      \do\u\do\v\do\w\do\x\do\y\do\z\do\A\do\B\do\C\do\D%
      \do\E\do\F\do\G\do\H\do\I\do\J\do\K\do\L\do\M\do\N%
      \do\O\do\P\do\Q\do\R\do\S\do\T\do\U\do\V\do\W\do\X%
      \do\Y\do\Z}
\def\UrlDigits{\do\1\do\2\do\3\do\4\do\5\do\6\do\7\do\8\do\9\do\0}
\g@addto@macro{\UrlBreaks}{\UrlOrds}
\g@addto@macro{\UrlBreaks}{\UrlAlphabet}
\g@addto@macro{\UrlBreaks}{\UrlDigits}
\makeatother
\titleformat*{\section}{\large\bfseries}
\titleformat*{\subsection}{\normalsize\bfseries}
\titleformat*{\subsubsection}{\normalsize\bfseries}
\usepackage[paperwidth=8.5in, paperheight=11in,margin=1 in]{geometry}
\usepackage{color}
\usepackage[dvipsnames]{xcolor}
\usepackage{tikz}

\DeclareRobustCommand{\blackline}{\raisebox{2pt}{\tikz{\draw[-,black!40!Black,solid,line width = 0.9pt](0,0) -- (5mm,0);}}}
\DeclareRobustCommand{\blueline}{\raisebox{2pt}{\tikz{\draw[RoyalBlue,solid,line width = 0.9pt](0,0) -- (5mm,0);}}}
\DeclareRobustCommand{\orangeline}{\raisebox{2pt}{\tikz{\draw[BurntOrange,solid,line width = 0.9pt](0,0) -- (5mm,0);}}}
\DeclareRobustCommand{\greenline}{\raisebox{2pt}{\tikz{\draw[Green,solid,line width = 0.9pt](0,0) -- (5mm,0);}}}
\DeclareRobustCommand{\redline}{\raisebox{2pt}{\tikz{\draw[Red,solid,line width = 0.9pt](0,0) -- (5mm,0);}}}

\DeclareRobustCommand{\blackdashedline}{\raisebox{2pt}{\tikz{\draw[-,black!40!black,dashed,line width = 0.9pt](0,0) -- (5mm,0);}}}

\DeclareRobustCommand{\tikzcircle}[2][red,fill=red]{\tikz[baseline=-0.5ex]\draw[#1,radius=#2] (0,0) circle ;}

\title{\fontsize{18}{36}\selectfont \textbf{Effects of varying initial conditions of ship encountering \\ 
\vspace{-3mm}
wave groups in computing extreme motion statistics}}
\author{\fontsize{14}{36}\selectfont Xianliang Gong, Yulin Pan \\
\fontsize{14}{36}\selectfont (Department of Naval Architecture and Marine Engineering, \\
\fontsize{14}{36}\selectfont University of Michigan, USA)}

\date{}
\usepackage{fancyhdr}

\begin{document}
\maketitle
\thispagestyle{fancy}
\begin{multicols}{2}

\section*{\selectfont ABSTRACT}
In computing ship motion statistics (e.g., exceeding probability) in an irregular wave field, it is a common practice to represent the irregular waves by a large number of wave groups and compute the motion statistics from the distribution of these wave groups. While this procedure significantly reduces the computational cost, the uncertainties introduced in this reduced-order computation have not been quantified. In general, the representation of a continuous wave field by separated groups loses information about wave phases, frequency modulation and initial conditions of ship when encountering the wave groups, among which the last one is arguably the most influential factor for the ship motion statistics. In this paper, we test three methods to incorporate the ship initial conditions into the computation of roll motion exceeding probability, namely the methods of \textit{natural initial condition},  \textit{pre-computed initial condition} and \textit{constant initial condition}. For different input wave spectra and ship motion dynamics (in terms of the parametric excitation level modeled by a nonlinear roll equation), the performances of the three methods are carefully benchmarked and the superiority of the \textit{natural initial condition} method is demonstrated. We finally show that the computation using the \textit{natural initial condition} method can be greatly accelerated through a sequential sampling algorithm making use of the variational heteroscedastic Gaussian process regression.
\setlength{\parindent}{0.5in}

\section*{INTRODUCTION}
Extreme ship motions can be triggered from various mechanisms (e.g., large waves, parametric rolls) for ships travelling in an irregular wave field. Although these motions occur with a low probability, they may cause severe damage to ships especially at high sea states. A reliable quantification of the extreme motion statistics, e.g., probability of ship motion exceeding a certain threshold, is of vital importance for assessment of the system reliability and reduction of the failure probability in the ship design process.

The computation of the motion exceeding probability, however, is a non-trivial task. Considering the high dimensions of the wavefield (as input to the ship-motion computation), the rareness of the extreme motions, and the high computational cost in CFD simulations of ship motion in waves, the direct Monte Carlo sampling in the whole input space (or a long simulation covering all wave conditions) is computationally prohibitive. Many methods have been developed to reduce the computational cost of the problem, with one critical effort to reduce the dimension of the input irregular wave field through a wave group representation \cite[e.g.][]{anastopoulos2017evaluation, cousins2016reduced}. Under such a representation (specifically the second reference) only two parameters of the group height $A$ and group length $L$ (with a given probability distribution) are needed to describe the wave field. Although this procedure allows the computation to be done with a much lower cost, certain information has been lost in using separate wave groups to represent a continuous wave field, namely the wave phases, the frequency modulation (of waves in a group) and the initial condition of the ship when encountering a wave group. The uncertainties associated with this lost information are usually ignored (or insufficiently quantified) in most existing works adopting the method of wave group parameterization.

Another critical development in accelerating the computation of motion exceeding probability is to use sequential sampling (or sequential Bayesian experimental design) method \cite[e.g.][]{echard2011ak,hu2016global,mohamad2018sequential,gong2021full} to reduce the Monte-Carlo sampling numbers required for the computation. The principle of these methods is to establish a Gaussian process as a surrogate model through the existing samples (and their responses), and optimize for the location of the next sample in the $(A,L)$ group parameter space. In applying these methods to naval platforms, the problem associated with the initial conditions (as well as other lost information) was usually ignored or somehow avoided by considering a static platform \citep{mohamad2018sequential}. In order to incorporate the effect of varying initial conditions into the sequential sampling framework, the authors of this paper have developed a new approach (VHGP-Seq) in \cite{gong2022sequential}, which uses the variational heteroscedastic Gaussian process as a surrogate model (thus quantifying the varying uncertainty in the responses from different groups) and solves a new optimization problem to locate the next-best sample. The VHGP-Seq method has been validated successfully in one case of computing the exceeding probability of roll motions, but this is not sufficient to guarantee the effectiveness of the method for wide ranges of the input wave spectra and ship roll dynamics.

The purposes of the present paper are twofold. In the first part, we test the performance of three existing methods to handle the initial conditions of ship encountering wave groups in computing the motion exceeding probability. Specifically, the first method \citep{anastopoulos2017evaluation, gong2021full} uses constant (usually zero displacement and velocity) initial condition of the ship motion for each wave group and therefore neglect the variation of the response induced from the variation of initial conditions (hereafter named \textit{constant initial condition}). The second method \citep{anastopoulos2019evaluation, silva2021towards} generates a distribution of initial conditions prior to the computation of statistics through simulation of ship motion in the same wave field (hereafter named \textit{pre-computed initial condition}). The third method \citep{gong2022sequential} simulates the ship motion from several groups ahead of the objective group and thus incorporates the initial condition in a natural way (hereafter named \textit{natural initial condition}). The three methods are applied to cases with different input wave spectra, motion thresholds and ship motion dynamics (in terms of the level of parametric excitation simulated by a nonlinear roll equation), and are compared to the true solutions generated as the ship goes through a sufficiently long continuous wave field (or time series). We demonstrate the superiority of the method of \textit{natural initial condition} in term of its agreement to the true solution in all tested cases.

In the second part of the paper, we test the performance of the VHGP-Seq sampling method in reproducing the true solution of the exceeding probability (or solution from the method of \textit{natural initial condition}) for a number of cases (as a representative subset of all cases used to test the methods of initial conditions). For all these cases, we show that the exceeding probability from the VHGP-Seq method converges to the true solution in $O(100)$ samples, instead of more than $O(1000)$ samples required in the random sampling approach.

\section*{METHOD}
\begin{center}
     \centering
     \includegraphics[width=\linewidth]{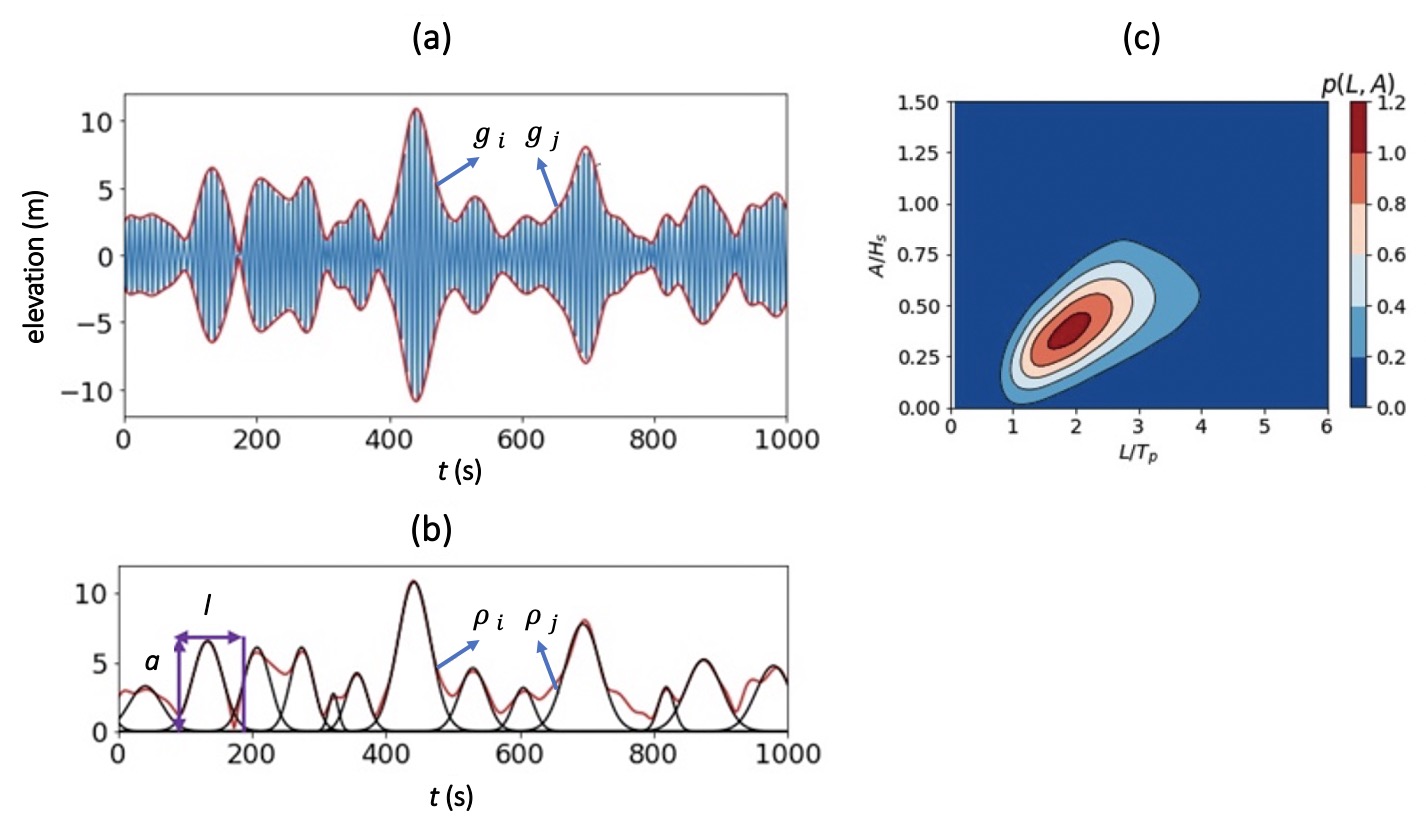}
     \captionof{figure}{(a) Wave field containing successive wave groups with surface elevation $\eta(t)$ (\blueline) and envelop $\rho(t)$ (\redline). (b) $\rho(t)$ (\redline) fitted by Gaussian wave groups $\rho_i(t)$ (\blackline). (c) The joint PDF of $L$ and $A$ computed from a large number of detected groups.}
     \label{fig: schematic}
\end{center}
We start from a narrow-band wave field with a sufficiently long time series $\eta(t)$, which contains successive wave groups $\{g_1, g_2, \cdots, g_n \}$ (see illustration from figure 1(a), and mathematical definitions shortly). Our objective is to compute the exceeding probability (probability of motion larger than a threshold, say $\theta_t$) when a ship goes through the wave field described by $\eta(t)$. In particular, we define the ground truth of the group-based exceeding probability as
\begin{equation}
    P_{true}  = \sum_{i=1}^{n} I(r_i) / n,
    \label{eq:ptrue}
\end{equation}
where $r_i$ is the maximum ship motion within a segment of $\eta(t)$ corresponding to the group $g_i$, simulated as the ship continuously goes through $\eta(t)$, $n$ is the total number of wave groups in $\eta(t)$, and $I$ is an indicator function:
$$
I(r_i) = \left\{
\begin{aligned}
    & 1         & 
    & if \; r_i > \theta_t                        \\
    & 0                              & 
    & if \; r_i \leq \theta_t  
\end{aligned}
\right..
$$
The computation of $P_{true}$, as defined in \eqref{eq:ptrue}, involves a simulation of the ship response from $\eta(t)$, which needs to be extremely long to cover (many times) all wave conditions associated with a given wave spectrum. When coupled with sophisticated motion simulations such as CFD, the computational cost can become prohibitively high. 

In order to reduce the overall computational cost, one may seek to parameterize the time series $\eta(t)$ by features of the wave groups and their associated probability distribution. There are two approaches to achieve this goal, the first by assuming a Markov-chain property of $\{g_i\}_{i=1}^{n}$ and computing the group distribution from the transitional probability following \cite{kimura1980statistical} and \cite{longuet1984statistical}, and the second by constructing the groups directly from the envelope process of $\eta(t)$ \citep{cousins2016reduced}. We follow the second approach in the current paper, where we compute the envelope process $\rho(t)$ from $\eta(t)$ through the Hilbert transform \citep{shum1984estimates}, and then construct Gaussian-like wave groups $\rho_i(t)$ which best fit $\rho(t)$ locally: 
\begin{equation}
    \rho_i(t)\sim a_i \exp \frac{-(t-t_i)^2}{2l_i^2},
    \label{Gaussian}
\end{equation}
where $t_i$, $a_i$, and $l_i$ are respectively  the  temporal  location, amplitude and length of group $g_i$ (see figure 1(b)). This construction relies on a group detection algorithm proposed by \cite{cousins2016reduced} and later improved by \cite{gong2021full}. Based on the detected groups, we can construct a two-dimensional description of the wave field with probability distribution $p_{A,L}(a,l)$ (figure 1(c)), and then compute an estimation of the group-based exceeding probability as
\begin{equation}
    P_{est}= \int I(r(a,l)) p_{A,L}(a,l) \mathrm{d} a \mathrm{d} l,
\label{ep_deter}
\end{equation}
where $r(a,l)$ is the maximum ship response in a group with parameter $(a,l)$ (with choice of initial conditions that will be described later). While \eqref{ep_deter} provides a reduced-order computation relative to \eqref{eq:ptrue}, the uncertainties associated with this simplification have not been fully quantified. Specifically, in using separated $(a,l)$ groups to represent $\eta(t)$, the uncertainties can come from certain lost information, including the wave phases in the group, frequency modulation in the group and the initial condition of the ship encountering a group. Among these factors, the ship initial condition is arguably (as will also be confirmed in our results) the most influential one for the motion exceeding probability. In the next three sections, we review different existing approaches to handle the encountering initial conditions, followed by a brief review of the VHGP-Seq sampling method and a presentation of our test framework.
\begin{figure*}
     \centering
     \includegraphics[width=16cm]{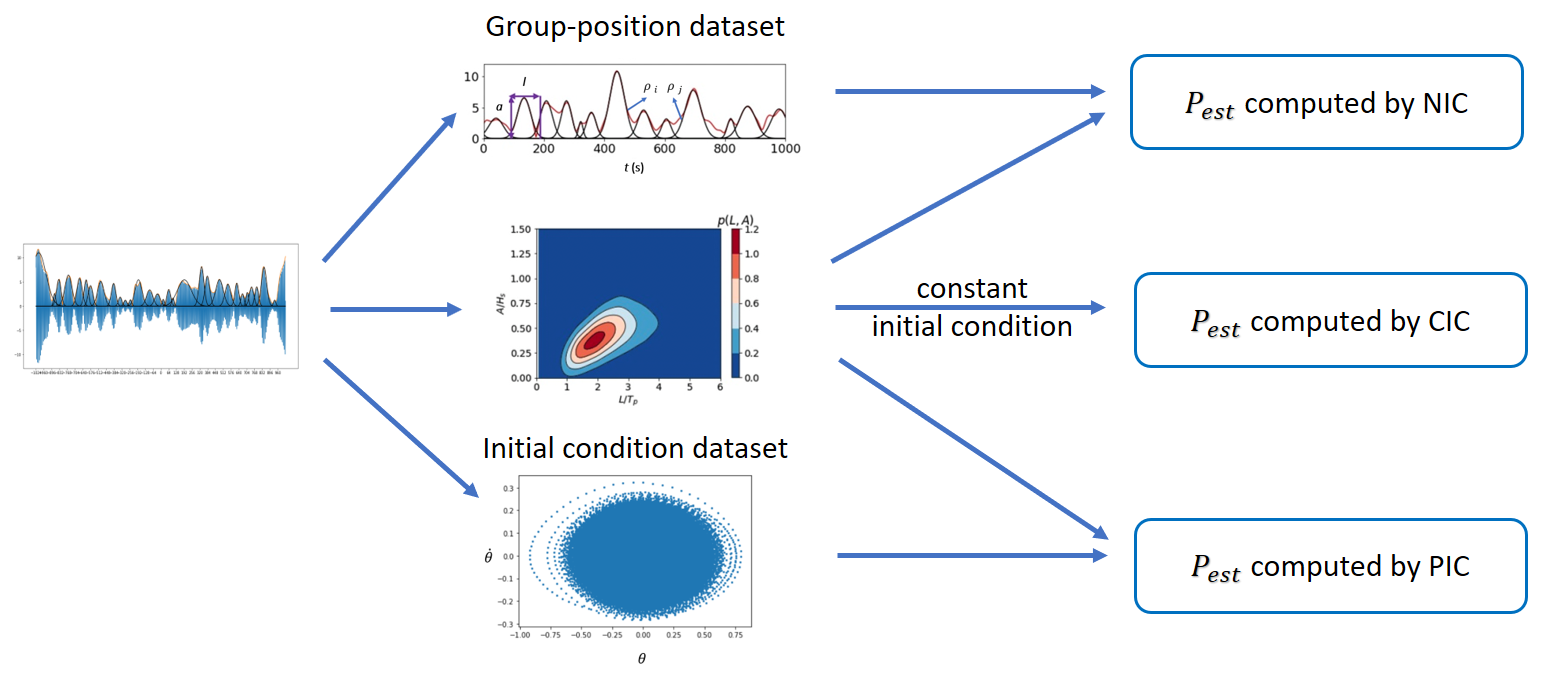}
     \caption{Flowchart of the three methods to incorporate the effects of initial conditions.}
     \label{fig: schematic}
\end{figure*}
\subsection*{Approaches for initial conditions}
We summarize three existing approaches to handle the initial conditions associated with $r(a,l)$.
\begin{enumerate}
   \item \textit{constant initial condition (CIC)}: We compute $r(a,l)$ for all groups using a constant initial condition (usually with zero displacement and velocity), and compute the exceeding probability directly by \eqref{ep_deter}. 

   \item \textit{pre-computed initial condition (PIC)}: We generate a distribution of ship motions (say, roll motion $\theta$ and velocity $\dot{\theta}$) from a simulation covering part of $\eta(t)$, which is then used as an ensemble of initial conditions to compute the response from an $(a,l)$ group. In this case, the response becomes a random variable denoted by $R(a,l)$, and accordingly we compute the exceeding probability by 
\begin{equation}
    P_{est} = \int \mathbb{P}(R(a,l)> \theta_t) p_{A,L}(a,l) \mathrm{d} a \mathrm{d} l.
\label{ep_rand}
\end{equation}
    \item \textit{natural initial condition (NIC)}: For a selected pair of parameters $(a,l)$, we simulate the responses from a number of detected $(a,l)$ groups. For each of these $(a,l)$ groups, we start the simulation from two or three groups ahead of the objective group to naturally capture the initial condition. The quantity $\mathbb{P}(R(a,l)> \theta_t)$ can then be approximated by the fraction of the responses that exceed $\theta_t$. Finally the exceeding probability is computed from \eqref{ep_rand}. This approach is in essence the same as a Monte-Carlo sampling of random groups from $\eta(t)$ with applications of \textit{NIC} to simulate the response from each of them.
\end{enumerate}

A sketch of these three approaches are further shown in figure 2. In application of all these methods, a necessary procedure is to recover a time series of surface elevation $\eta_i(t)$ from a given $(a,l)$ group $g_i(t)$ (as the input to simulate the response). In this work, the time series $\eta_i(t)$ is obtained by 
\begin{equation}
   \eta_i(t)=\rho_i(t) \cos(\omega_p t),
   \label{etai}
\end{equation}
where $\omega_p$ is the peak angular frequency of the wave spectrum under consideration, $\rho_i(t)$ is connected to $a$ and $l$ from \eqref{Gaussian}. It is clear from \eqref{etai} that the information about wave phases (and frequency modulation) of the original series $\eta(t)$ is lost in all three methods. Other than this lost information, the \textit{NIC} method seems to be the most sophisticated one to incorporate the effect of initial conditions. Therefore, it is not unreasonable to expect that the method of \textit{NIC} may provide the closest result to the true solution \eqref{eq:ptrue}. In fact, A direct comparison between the result from \textit{NIC} and the true solution \eqref{eq:ptrue} will reveal whether the ship initial condition is the dominant factor of uncertainty and whether the proposed method is sufficient to capture its effect on the exceeding probability.

We also remark that direct applications of the \textit{PIC} and \textit{NIC} methods (i.e., with random sampling in the $(A,L)$ parameter space) to compute the exceeding probability are still relatively expensive. Taking the \textit{NIC} method as an example, while it is possible to explore the reduction of the computational cost by simulating only a subset of wave groups out of those in \eqref{eq:ptrue}, this reduction may not be sufficient. It should be noted here that our first focus of the paper is to understand the accuracy of the three methods in incorporating the effect of initial conditions, irrespective of the computational cost. The reduction of the computational cost will further be explored in the context of VHGP-Seq method that is next discussed.

\subsection*{VHGP-Seq sampling method}
We next briefly review the VHGP-Seq sampling method, that can be used to significantly reduce the number of samples required in the computation using the \textit{NIC} method. The principle of the VHGP-Seq is to first establish a surrogate model of the stochastic response $R(a,l)$ using the variational heteroscedastic Gaussian process (VHGP) from existing samples. In this surrogate model, the random variable $R(a,l)$ (for given $a$ and $l$) is approximated by a normal distribution, and two Gaussian processes are used to model the mean and variance of $R(a,l)$. Then the next-best sample location is chosen based on an acquisition optimization problem which maximizes the contribution of the sample to the computation in \eqref{ep_rand}.  

Since the variational algorithm is used in VHGP, the surrogate model can be established in an efficient manner, with a computational cost that is only twice of a standard Gaussian process regression. We remark that the VHGP-Seq method is so far the only available method for sequential sampling in problems with a stochastic response function whose standard deviation varies with the input. The \textit{NIC} problem is clearly an example of this type of problems that are generally encountered in many other contexts. For more details of VHGP-Seq, we refer the readers to our previous paper \citep{gong2022sequential}.

\subsection*{Test framework}
Our purpose is to benchmark the performances of \textit{CIC}, \textit{PIC} and \textit{NIC} in computing the exceeding probability, as well as the performance of VHGP-Seq in accelerating the computation. Since the true solution \eqref{eq:ptrue} is needed in the comparisons, computation of the ship motion using CFD becomes impractical because of the prohibitively high computational cost. In this work, we instead employ a nonlinear roll equation \citep{umeda2004nonlinear} to calculate the time series of ship roll $\theta(t)$ from the wave input $\eta(t)$:
\begin{align}
    \ddot{\theta}+\alpha_1 \dot{\theta}+\alpha_2 \dot{\theta}^3 +&(\beta_1+\epsilon_1 \eta(t) )\theta+\beta_2 \theta^3 \nonumber\\
    &= \epsilon_2 \eta(t).
\label{roll}
\end{align}
The equation \eqref{roll} phenomenologically models the ship roll motion in irregular waves incorporating nonlinear resonance and parametric roll mechanisms. We use empirical parameters $\alpha_1=0.01$, $\alpha_2=0.052$, $\beta_1=0.1175$, $\beta_2=-0.09$, and $\epsilon_2=0.004$. The parameter $\epsilon_1$ is varied to model different levels of parametric excitation, as one important mechanism leading to large responses, with its value chosen from $\epsilon_1 =\{0, 0.02, 0.04\}$. Given input $\eta(t)$ and initial condition, \eqref{roll} is numerically integrated with a 4th-order Runge-Kutta method to calculate the roll motion $\theta(t)$.

In addition, we test all methods for four different input wave spectra. The first three are drawn from spectra of a Gaussian form 
\begin{equation}
F(k)\sim \exp \frac{-(k-k_0)^2}{2\mathcal{K}^2},  
\label{spectrum}
\end{equation}
with significant wave height $H_s=12m$, peak (carrier) wavenumber $k_0=0.018 m^{-1}$ (corresponding to peak period $T_p=15s$). The parameter $\mathcal{K}$, which models the bandwidth of the spectrum, is chosen from $\mathcal{K} = \{0.02, 0.03, 0.04\} m^{-1}$ representing spectra of increasing bandwidth (although all of them are relatively narrow-banded). We term the three spectra with increasing $\mathcal{K}$ as S1, S2, and S3 hereafter. The fourth spectrum, termed S4, is in the JONSWAP form with parameters $H_s=12m$, $T_p=15s$ and $\gamma=6$. Compared to S3, S4 has similar peak-mode region but a longer tail in power-law form (instead of exponential tail in S3).

The true solutions $P_{true}$ in all cases are computed by \eqref{eq:ptrue} from a time series $\eta(t)$ of eight million $T_p$. To compute $P_{est}$ using $\textit{CIC}$, we use constant initial conditions of $\theta=\dot{\theta}=0$ for all wave groups in simulating \eqref{roll}. For $\textit{PIC}$, we establish the distribution of initial conditions using simulation results for thirty thousands $T_p$. For $\textit{NIC}$, we simulate \eqref{roll} starting from two wave groups ahead of the objective $(a,l)$ wave group. We make sure that sufficient number of random samples are used for convergent solutions in the computations applying $\textit{CIC}$, $\textit{PIC}$ and $\textit{NIC}$.

\section*{RESULTS}

\subsection*{\textit{CIC}, \textit{PIC}, and \textit{NIC} for initial conditions}

\begin{figure*}
     \centering
     \begin{minipage}[b]{\linewidth}         
         \includegraphics[width=\linewidth]{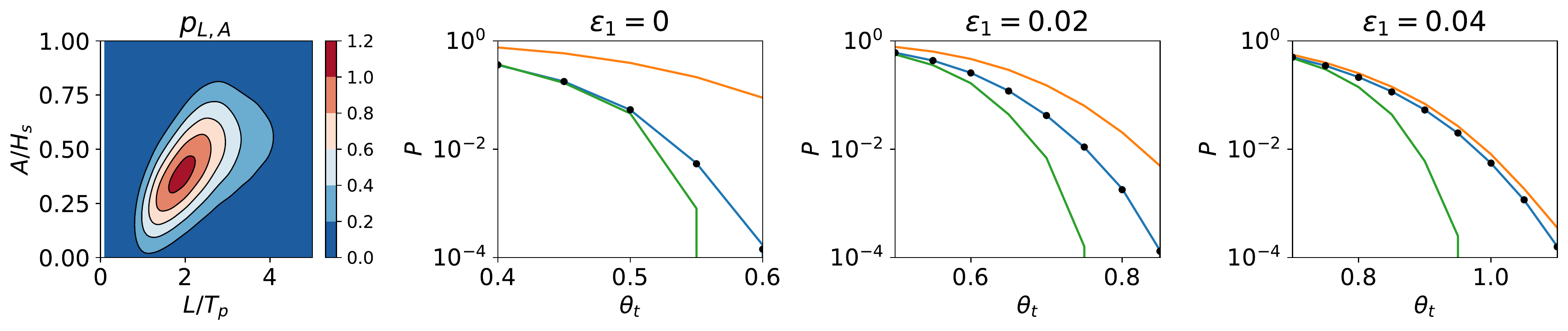}
         \caption*{(a) Gaussian Spectrum S1 with $\mathcal{K}=0.02 m^{-1}$ \break}
     \end{minipage}
     \begin{minipage}[b]{\linewidth}         
         \includegraphics[width=\linewidth]{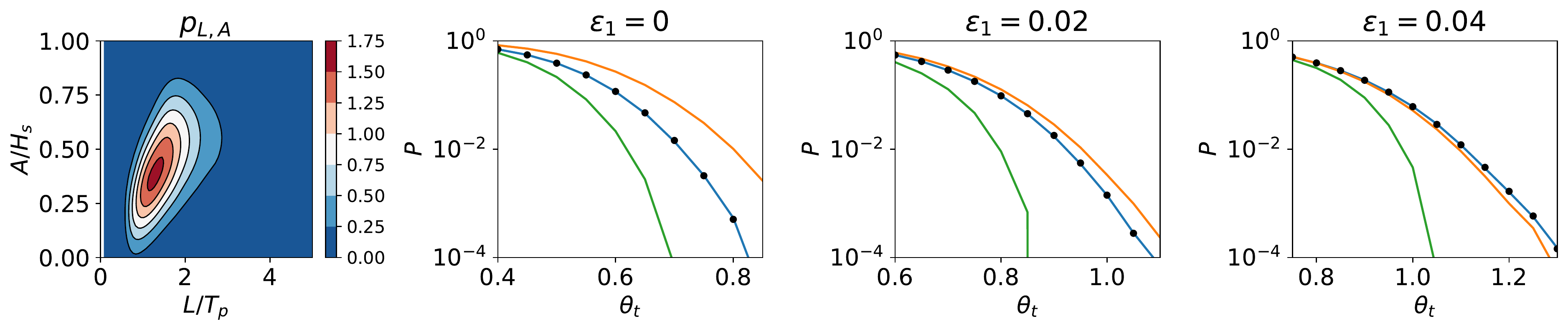}
         \caption*{(b) Gaussian Spectrum S2 with $\mathcal{K}=0.03 m^{-1}$ \break}
     \end{minipage}
     \begin{minipage}[b]{\linewidth}         
         \includegraphics[width=\linewidth]{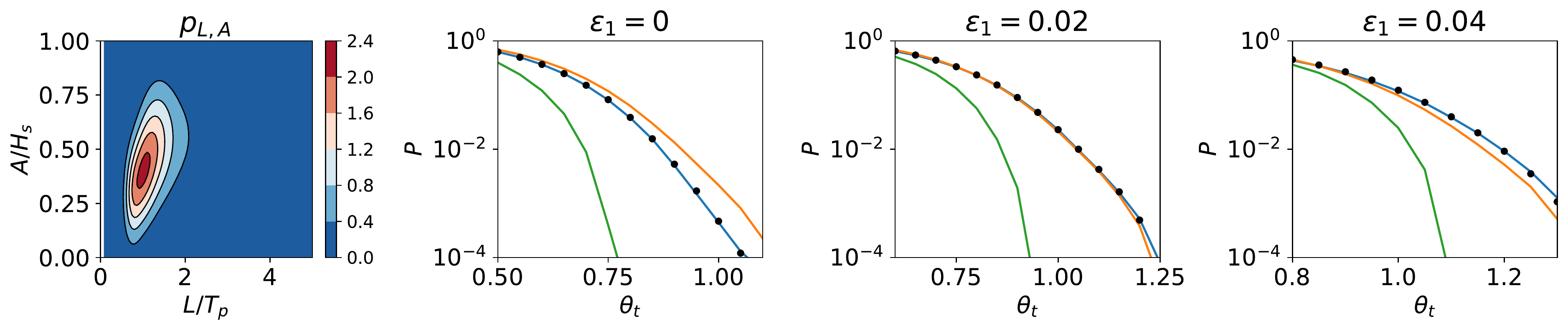}
         \caption*{(c) Gaussian Spectrum S3 with $\mathcal{K}=0.04 m^{-1}$ \break}
     \end{minipage}
     \begin{minipage}[b]{\linewidth}         
         \includegraphics[width=\linewidth]{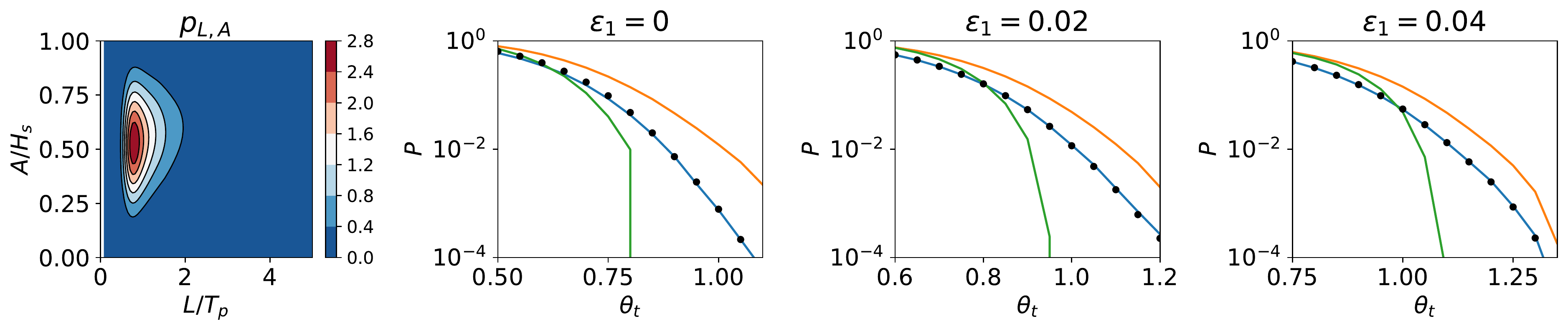}
         \caption*{(d) Jonswap Spectrum S4 with $\gamma=6$}
     \end{minipage}
        \caption{First column: the joint PDFs of $(A,L)$ obtained from wave fields described by different wave spectra. Second to forth columns: $P_{est}$ as a function of threshold $\theta_t$ (in radians) computed by \textit{PIC} method (\orangeline), \textit{NIC} method (\blueline) and \textit{CIC} method (\greenline), as well as true solution $P_{true}$ (\tikzcircle{1pt, Black}) for $\epsilon_1$=\;0, 0.02 and 0.04 from left to right, and different spectra from top to bottom.}
        \label{fig: exceeding}
\end{figure*}

We first benchmark the performances of \textit{CIC}, \textit{PIC}, and \textit{NIC} methods in computing $P_{est}$. In figure \ref{fig: exceeding} we plot the true solution $P_{true}$, as well as $P_{est}$ from \textit{CIC}, \textit{PIC}, and \textit{NIC} methods as functions of the threshold $\theta_t$, for different spectra S1$\sim$S4 and parametric excitation level $\epsilon_1$. For the same spectrum and a given $\theta_t$, $P_{true}$ increases with the increase of $\epsilon_1$ (by examining figures in the same row). This behavior signifies the effect of parametric roll in generating large responses. 

Comparing to $P_{true}$, the \textit{CIC} method significantly underestimates the exceeding probability for large motion thresholds (i.e., $P_{est} \ll P_{true}$ for large $\theta_t$) for all cases. The reason for these deviations lies in the fact that some large motions are induced by non-zero initial conditions when the ship encounters a wave group, and these scenarios are completely ignored when applying the \textit{CIC} method. For example, the parametric roll phenomenon may take a few wave groups to develop. If the initial condition for the target group (where the large response from parametric roll is supposed to happen) is artificially set to zero, the large response is then erroneously missed in the \textit{CIC} method.  

Relative to \textit{CIC}, the \textit{PIC} method shows a better performance in terms of its better agreement with $P_{true}$ in most cases. We also see that the performance of \textit{PIC} is better for larger values of $\epsilon_1$, i.e., higher level of parametric excitation. This behavior can be argued from the following reasoning: In principle, the \textit{PIC} method tends to provide an overestimation of initial conditions because its distribution of initial conditions is drawn from a time series $\theta(t)$ which includes both those at the starting point of and within the group (with the latter resulting in an overestimation). This tendency of overestimation somehow accounts for the increased probability of parametric roll with the increase of $\epsilon_1$, resulting in a favorable estimation seen in figure \ref{fig: exceeding}. 

Finally, the results from the \textit{NIC} method closely follow $P_{true}$ in all cases (hardly distinguishable with a log axis). We recall that potential deviation of the \textit{NIC} method from the computation \eqref{eq:ptrue} can be caused by the neglect of wave phases and frequency modulation of $\eta(t)$ in  \textit{NIC}, and that the  \textit{NIC} attempts to trigger an initial condition by simulating the previous two groups (which cannot be validated beforehand). The favorable agreement achieved in figure \ref{fig: exceeding} indicates important understanding on both aspects: (i) the effects of wave phases and frequency modulation in the wave group are indeed not significant to the exceeding probability (or they are of secondary importance compared to that of varying initial conditions). (ii) the procedure in the \textit{NIC} method is sufficient to capture the correct initial condition for the ship encountering a given wave group.

\begin{figure*}
     \centering
     \includegraphics[width=14cm]{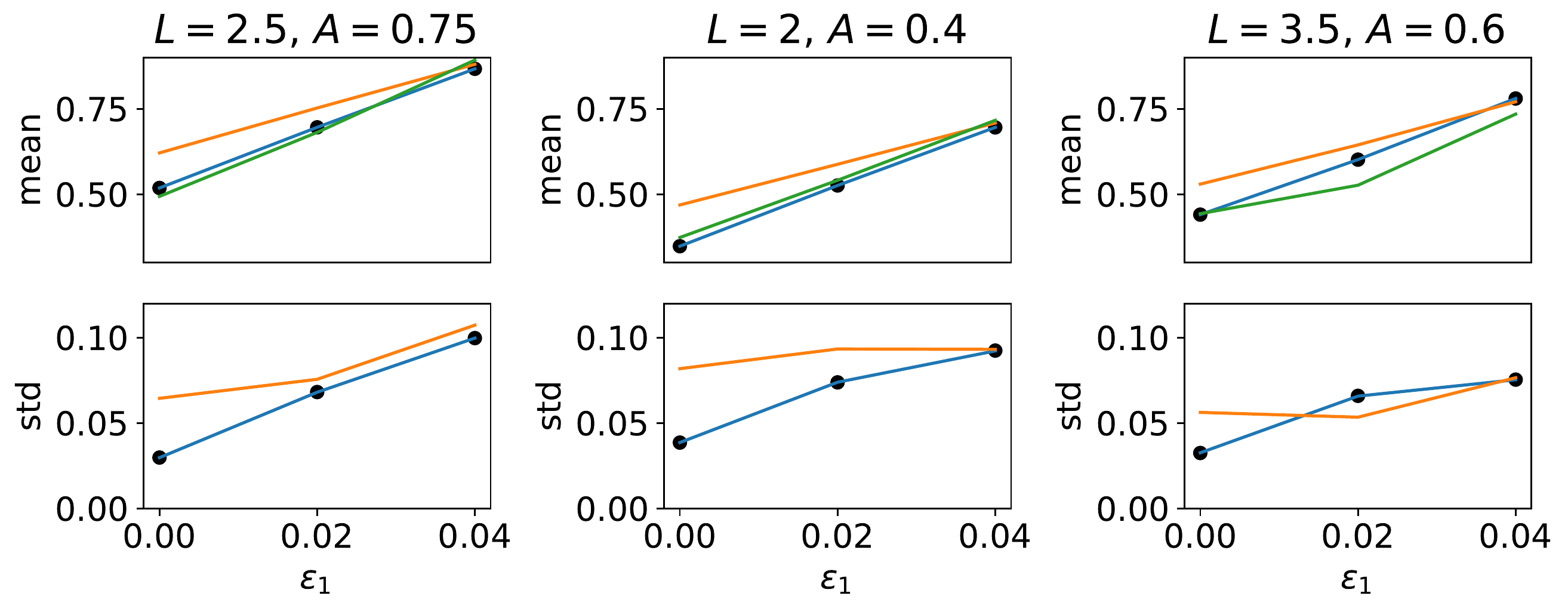}
     \captionof{figure}{(upper panel) mean and (lower panel) standard deviation of $R(a,l)$ for three choices of $(a,l)$ parameters for the S1 spectrum, computed by \textit{PIC} (\orangeline), \textit{NIC} (\blueline), \textit{CIC} (\greenline) and the ground truth (\tikzcircle{1pt, Black}).}
        \label{fig: uncertainty}
\end{figure*}
The performances of the three methods can be further elucidated by examining their computations regarding wave groups sharing the same $(a,l)$ parameter. In figure \ref{fig: uncertainty}, the responses $R(a,l)$ from three selected $(a,l)$ parameters are plotted, in terms of the mean and standard deviation of $R(a,l)$ from the three methods and the true solutions. It can be seen that the \textit{CIC} method, by applying zero initial conditions, well captures the mean of $R(a,l)$ but ignores the standard deviation, i.e., the variation in the ensemble of $R(a,l)$. Under this situation, it can be expected that some initial conditions leading to exceeding motions are neglected in \textit{CIC}, resulting in its underestimation of the exceeding probability. The \textit{PIC} method, on the other hand, provides overestimation of both the mean and standard deviation of $R(a,l)$ for smaller values of $\epsilon_1$, but relatively accurate estimation at larger $\epsilon_1$. This explains the performance of \textit{PIC} observed in figure \ref{fig: exceeding}. Finally, the \textit{NIC} method captures accurately both the mean and standard deviation of $R(a,l)$ for chosen $(a,l)$ parameters. 

We remark that the above conclusions are drawn based on \eqref{roll} with a set of selected parameters. To generalize the conclusion with high confidence, it is desired to test the results for more parameter values in \eqref{roll}. However, considering that our current set of parameters (and the equation) are not chosen in a specific way, one can anticipate that the conclusion holds for much broader situations of sea conditions and ship motion dynamics.

\subsection*{VHGP-Seq to accelerate the \textit{NIC} method}
While the \textit{NIC} method provides very accurate estimation of the exceeding probability, a direct application with random sampling to compute \eqref{ep_rand} can be prohibitively expensive if CFD is used to obtain the motion for given $a$ and $l$. For example, the results in figure \ref{fig: exceeding} are obtained with 200,000 random samples of wave groups.  

To accelerate the \textit{NIC} method, we apply our recently developed VHGP-Seq sampling method to reduce the number of required samples (for an accurate estimation of the exceeding probability). Figure \ref{fig: vhgp} shows a typical example of $P_{est}$ computed by the \textit{NIC} method with random and VHGP-Seq samplings for the S2 spectrum, $\theta_t=0.9$ and $\epsilon_1=0.02$. In particular, we include the mean and standard deviation of $P_{est}$ computed by 100 realizations of both methods. For \textit{NIC} with random sampling, the mean of $P_{est}$ always lies in the vicinity of $P_{true}$ due to the basic properties of Monte-Carlo random sampling (i.e., the Monte-Carlo random sampling is an unbiased estimator from law of large numbers). However, to obtain meaningful solution in one realization we require the standard deviation of $P_{est}$ to be reduced to a low level, i.e., the result converges to the true solution in a single experiment. From figure \ref{fig: vhgp} it is clear that it takes random sampling method more than O(1000) samples for the standard deviation of $P_{est}$ to reach a satisfactory state for estimation. In contrast, when VHGP-Seq sampling is applied, less than O(100) samples are required for the convergence of the estimated result to the true solution, i.e., the mean of $P_{est}$ agreeing with $P_{true}$ and the standard deviation of $P_{est}$ reduced to a low level.

\begin{center}
     \centering
     \includegraphics[width=6cm]{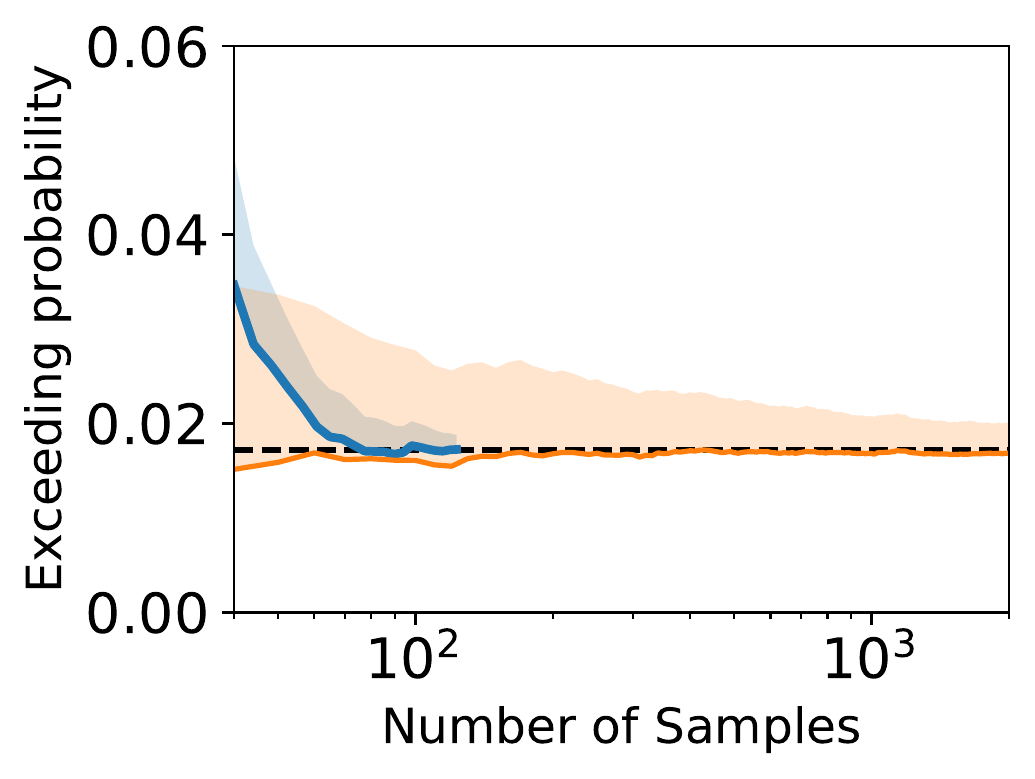}
     \captionof{figure}{The exceeding probability for the S2 spectrum, $\theta_t=0.9$ and $\epsilon_1=0.02$, with $P_{est}$ computed by the \textit{NIC} method with VHGP-Seq sampling (\blueline) and random sampling (\orangeline). The solid line and shaded region represent the mean and one-sided standard deviation of $P_{est}$ respectively, computed from 100 realizations of both methods. The true solution $P_{true}$ is indicated by (\blackdashedline).}
     \label{fig: vhgp}
\end{center}

Furthermore, we apply the VHGP-Seq sampling method to more cases selected from figure \eqref{fig: exceeding}. In Table 1 we show the results of the \textit{NIC} method with VHGP-Seq and random samplings for $\epsilon_1=0.02$ and different other selected parameters. Specifically, we present the mean and standard deviation of $P_{est}$ computed by 100 realizations of the random and VHGP-Seq samplings. In all cases, with 120 samples it is found that the VHGP-Seq sampling provides results with the means closer to $P_{true}$ and standard deviations much smaller than those from random sampling. With these evidences we conclude that O(100) samples are sufficient for the VHGP-Seq method to provide satisfactory estimation of exceeding probability of $O(10^{-3})$ when coupled with the \textit{NIC} method.

\begin{center}
\captionof{table}{Mean and standard deviation of $P_{est}$ at 120 samples of the VHGP-Seq and random sampling methods compared to the true solution $P_{true}$ for $\epsilon_1=0.02$ and other selected parameters. The standard deviation is expressed by its percentage of the mean.}
\begin{tabular}{|c|c|c|cl|}
\hline
\multirow{2}{*}{S} & \multirow{2}{*}{$\theta_t$} & \multirow{2}{*}{$P_{true}$} & \multicolumn{2}{c|}{mean (std) at 120 samples} \\ \cline{4-5} 
   &      &        & \multicolumn{1}{c|}{sequential}   & \multicolumn{1}{c|}{random} \\ \hline
S1 & 0.7  & 0.0421 & \multicolumn{1}{c|}{0.0416(5\%)}  & 0.0420(44\%)                \\ \hline
S2 & 0.9  & 0.0172 & \multicolumn{1}{c|}{0.0174(8\%)}  & 0.0156(67\%)                \\ \hline
S3 & 1    & 0.0232 & \multicolumn{1}{c|}{0.0233(8\%)}  & 0.0242(61\%)                \\ \hline
S4 & 1.05 & 0.0076 & \multicolumn{1}{c|}{0.0082(12\%)} & 0.0105(101\%)               \\ \hline
\end{tabular}
\end{center}

\section*{CONCLUSIONS AND DISCUSSION}
\begin{center}
     \centering
     \includegraphics[width=7.5cm]{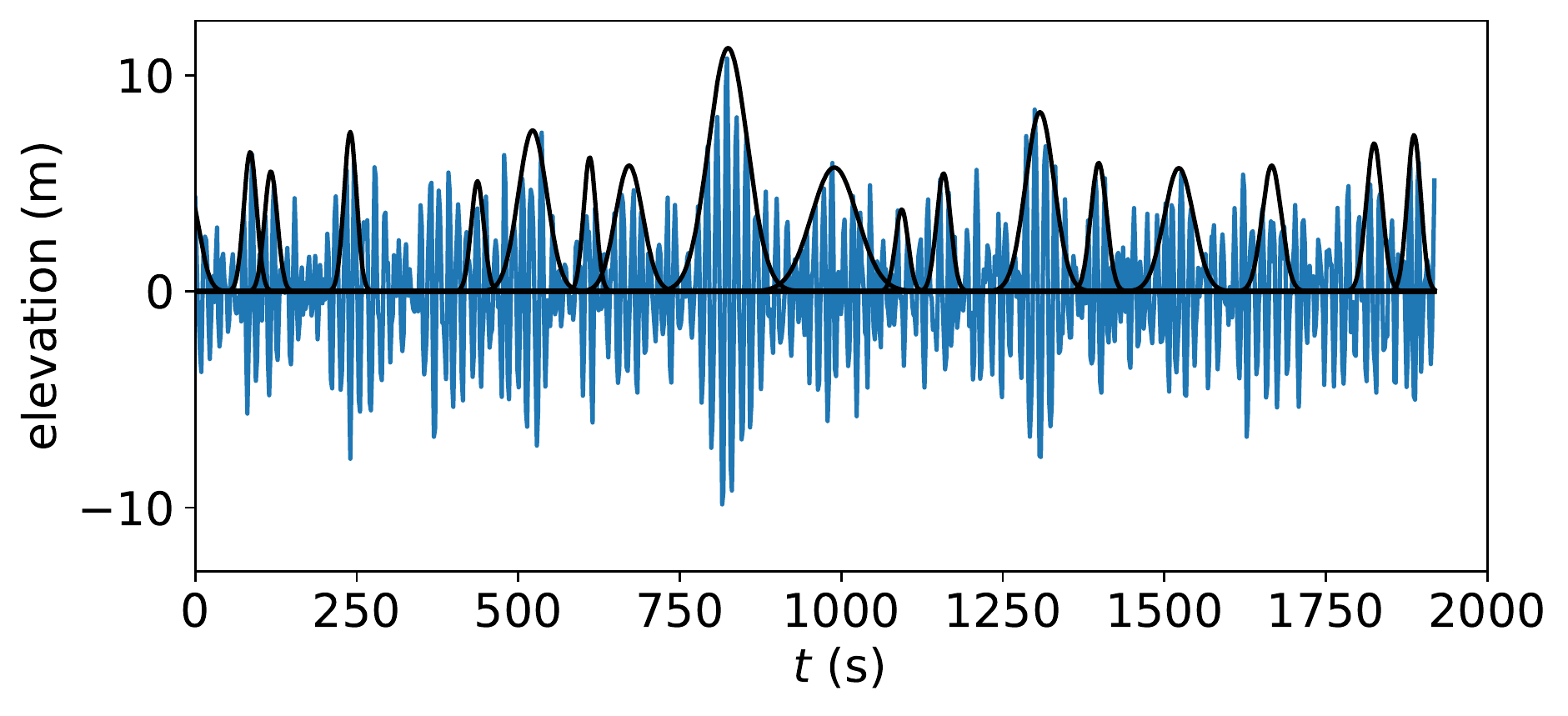}
     \captionof{figure}{$\eta(t)$ (\blueline) generated by JONSWAP spectrum and the detected Gaussian wave groups (\blackline).}
     \label{fig: jonswap}
\end{center}
In this paper, we investigate the uncertainties associated with the wave group representation of surface elevation series $\eta(t)$ in estimating the motion exceeding probability. We find that the uncertainties are mainly caused by the loss of initial conditions of the ship encountering wave groups when the reduced-order group representation is used. Three methods to handle the initial condition problem are tested, namely the methods of \textit{constant initial condition (CIC)}, \textit{pre-computed initial condition (PIC)} and \textit{natural initial condition (NIC)}. For a broad range of cases with different input wave spectra, levels of parametric excitation and motion thresholds, \textit{NIC} outperforms \textit{PIC} and \textit{CIC}, and provides accurate estimation of the exceeding probability in remarkable agreement with the true solution. Furthermore, we explore the application of the VHGP-Seq sampling method to accelerate the computation using the \textit{NIC} method, and show that only O(100) samples are needed to obtain converged result of $P_{est}$ of $O(10^{-3})$ instead of more than O(1000) samples in random sampling.

We finally remark on the group-based probability used in this paper. The exceeding probability \eqref{eq:ptrue} is always well defined given a group detection from $\eta(t)$, and can be computed with the methods presented in this paper. However, for certain cases the practical relevancy of \eqref{eq:ptrue} needs to be further discussed. This is especially for cases with an input wave spectrum consisting of many short wave components so that the wave group structure cannot be uniquely defined. An example is the JONSWAP spectrum used in this paper, which has a realistic power-law tail that decays much slower than the exponential tail in the Gaussian-form spectrum. Figure \ref{fig: jonswap} shows a segment of $\eta(t)$ described by the JONSWAP spectrum together with the detected groups. While these groups provide the basis for applying all calculations presented in this paper, the group-based exceeding probability may vary depending on which and how many groups (as the number of groups serves as the denominator in computing the probability) are detected from $\eta(t)$. This information can vary with the settings in the group detection algorithm, so that the group-based probability may not be intuitive to interpret in practice. 

In general, a more robust measure that we can apply for this type of problem is the temporal exceeding probability, i.e., ratio of the motion exceeding time and the total time. This measure remains (largely) an invariant even if the group-parameterization framework is applied for the computation. More specifically, as long as the groups that lead to large exceeding motions are detected (which is practically true), the results of temporal exceeding probability will remain a constant. However, the computation of temporal exceeding probability involves other issues to be resolved. We will provide detailed algorithm and results in an upcoming paper.

\section*{ACKNOWLEDGEMENT}
This research is supported by Office of Naval Research grant N00014-20-1-2096. We thank the program manager Dr. Woei-Min Lin for several helpful discussions on the research.

\bibliographystyle{plainnat}
\bibliography{mybib.bib}
 
\begin{appendices}

\end{appendices}

\end{multicols}
\end{document}